# Predicting Bank Loan Default with Extreme Gradient Boosting

Rising Odegua
Department of Computer Science Ambrose Alli University
Ekpoma, Edo state, Nigeria
risingodegua@gmail.com

**Abstract**
Loan default prediction is one of the most important and critical problems faced by banks and other financial institutions as it has a huge effect on profit. Although many traditional methods exist for mining information about a loan application, most of these methods seem to be underperforming as there have been reported increases in the number of bad loans. In this paper, we use an Extreme Gradient Boosting algorithm called XGBoost for loan default prediction. The prediction is based on a loan data from a leading bank taking into consideration data sets from both the loan application and the demographic of the applicant. We also present important evaluation metrics such as Accuracy, Recall, precision, F1-Score and ROC area of the analysis. This paper provides an effective basis for loan credit approval in order to identify risky customers from a large number of loan applications using predictive modeling.

## 1. Introduction

The credit lending business of the banking industry has seen rapid progress and extreme competitions from numerous credit startups. At the same time, the increase in credit application and consumption has also led to the increase in losses resulting from bad credits.

Credit loan refers to credits provided by banks or financial institutions in general to individuals/consumers which is payable at an agreed date with/without a specified interest.

Credit loans are usually provided for many purposes some of which are personal use, educational purposes, medical purposes, travelling and business purposes.

The increase in the application for loans plus the rapidly growing competition means financial institutions must build effective models that can capture the information in the available data, and create robust predictive models that can help minimize the chances of bad credit.

Through numerous modern predictive modeling, financial institutions can get insights into applicant's behavior, consumption patterns, default predictors and characteristics.

Numerous studies have been conducted in order to identify the important factors that can affect the loan repayment, these studies are important as they help to maximize profit for banks.

According to Manjeet et al (2018) there are seven types of variables that may influence consumer loan default; these are relationship between consumer and creditors, consumer's annual income, debt-income ratio, occupation, home ownership, work duration and whether or not consumer possesses a saving/checking account.

Chang, Cow, and Liu (2002) also indicated that an applicant's individual characteristics such as age, attitude could also influence risk behavior of borrowers.

In a work by Steenackers and Goovaerts (1989), the key factors that may influence loan default are borrower's age, location, resident/work duration, owner of phone, monthly income, loan duration, whether or not applicant works in a public sector, house ownership and loan numbers.

Another study by Ali Bangherpour (2015) on a large dataset within the period of 2001-2006 indicated that loan age was the most important factor when predicting loan default while market loan-to-value was the most effective factor for mortgage loan applications.

In addition to identifying factors that may influence loan default, there is also a need to build robust and effective machine learning models that can help capture important patterns in credit data. The choice of model is of great importance as the chosen model plays a crucial role in determining accuracy, precision and efficiency of a prediction system. Numerous models have been used for loan default prediction and although there is no one optimal model, some models definitely do better than others.

Manjeet et al (2018) used a Neural Network to predict loan default on data from a lending club bank and inferred that a Neural Network performs better than most traditional models.

Li Ying (2018) did a comparative study of three models (Random Forest, Logistic Regression and Support Vector Machines) on bank credit data and found out that Random Forest does better at the task.

Ali Bangherpour (2015) applied four algorithms to a large dataset of about 20 million observations and compared the result with a logistic regression model-a widely used traditional model for classification task.

Yang et al used a Logistic Regression model on a credit company's data to identify the most influential corporate financial indicators.

Lee et al (2006) conducted a study on data from a credit company in Taiwan, and compared different models such as CART, Neural Networks, Linear regression etc. The study concluded that the CART had better accuracy than the other models.

In this paper, we use a gradient boosting algorithm XGBoost to study and analyze bank loan dataset and suggest some of the important factors/variables that may influence loan repayment. We also present evaluation statistics (Accuracy, Precision, Recall, Confusion Matrix, f1-Score ad ROC area) of our model.

The remainder of the paper is organized as follows: section 2 provides a background and overview of Gradient Boosting and XGBoost. Section 3 explains the data collection, pre-processing steps and some basic description of the data. In Section 4 we discuss our results and findings and section 6 concludes.

## 2. Background

In this section, we describe the machine learning algorithm we use to forecast credit loan default in a supervised machine learning setting. In Supervised learning setting, a predictor/learner/estimator is presented with input-output pairs *{($x_1$, $y_1$), ($x_1$, $y_1$), ..., ($x_n$, $y_n$)}* for some function y = f(x). In this analysis, the supervised learning problem is posed as a classification problem since the target/output is composed of discrete binary variables

(good/bad credit). There exist many machine learning algorithms such as Logistic Regression, Neural Networks, Vector Machines, Decision Trees etc. that can be used for this classification task but we use a technique of ensembling called Gradient Boosting (XGBoost) as these have been shown to outperform many traditional algorithms in structured data settings like we have here.

### 2.1 Overview of Boosting

Boosting is a popular ensemble technique that originated in answer to a question posed by Kearns and Valiant (1989). That is, whether a "weak learner" could be made better by using some form of modification. This is similar to the statistical question of whether it is possible to create a "good hypothesis" from a "poor hypothesis". This was discovered to be possible by Schapire (1990) and the first boosting algorithm Adaptive Boosting (AdaBoost) was created by Freund and Schapire (1996).

The concept of boosting is to correct the mistakes made by earlier learners and improving on those areas Zhou (2012). I.e. we construct new base learners that are more correlated to the negative gradient of our objective function.

Boosting can also be seen as a kind of stage wise "additive modeling" (Buhlmann and Hothorn) in that it is an additive combination of a simple base estimator. Boosting is similar to bagging; the only difference is how they are trained.

Gradient Boosting (Friedman 2000) is a type of boosting where the objective is treated as an optimization problem and training is done using weight updates by gradient descent.

---

**Algorithm I** Friedman's Gradient Boosting Algorithm

---

**Inputs:**
- Input data (x, y) N i=1
- Number of iterations M
- Choice of the loss-function (y, f)
- Choice of the base-learner model h(x, θ)

**Algorithm**:
1: initialize $f_0$ with a constant
2: **for** $t$ = 1 to M do
3:   compute the negative gradient $g_t(x)$
4:   fit a new base-learner function h(x, $θ_t$)
5:   find the best gradient descent step-size $ρ_t$:

$$ρ_t = \arg\min_ρ \sum_{i=1}^{N} φ\,[yi, fi-1\,(xi) + ρh\,(xi, θt)]$$

6:   update the function estimate:
     $f_t \leftarrow f_t-1 + ρ_t h(x, θ_t)$

7: **end for**

### 2.2 XGBoost

XGBoost-Extreme Gradient Boosting (Chen and Guestrin 2016) is a scalable and highly efficient boosting system. It has been shown to achieve state-of-the-art results on many machine learning tasks. In XGBoost algorithm unlike the traditional gradient boosting, the process of adding weak learners does not happen sequentially; it approaches this phase in parallel using a multithreaded pattern, thereby resulting in proper utilization of hardware resources leading to greater speed and efficiency.

Some important features that make XGBoost more efficient than traditional boosting algorithms are:

1. Sparse aware implementation.
2. Weighted quantile sketch for approximate tree learning.
3. Cache-aware access.
4. Blocks for out-of-core computation.

## 3 Data

This paper used the loan credit data set provided by Data Science Nigeria and hosted on the Zindi platform. There are three sets of data:

1. Demographic data: This contains description and details about loan applicants. Attributes such as birthdates, addresses and account types can be found here. The data is made up of 4346 samples and 9 attribute features.
2. Performance data: This contains further details of each applicant and a feature (**good_bad_flag**) which describes if an applicant is a good or bad debtor. The data contains 4368 samples and 10 attributes.
3. Previous loan data: This contains all previous loan applications. The data contains 18183 samples and 12 attributes.

| Data | No of Attributes | Samples |
|---|---|---|
| Demographic | 9 | 4346 |
| Loan performance | 10 | 4368 |
| Previous Loan | 12 | 18183 |

**Table** 1. Description of data sets

### 3.1 Data Preparation

After data collection, we performed data processing and wrangling. The framework for preprocessing the data sets in Fig. 1.

## 4 Findings and Discussions

Based on the pre-processed data sets, we use the XGBoost Classifier algorithm implemented in python programming language. We trained our classifier on the clean data set using the feature **good_bad_flag** as target. We used a 5 fold cross validation to avoid recording a bias performance. Optimal parameters were obtained using a Grid Search (see Table 2). For evaluation, we use five metrics; Accuracy, F1-Score, Recall, Precision and ROC area.

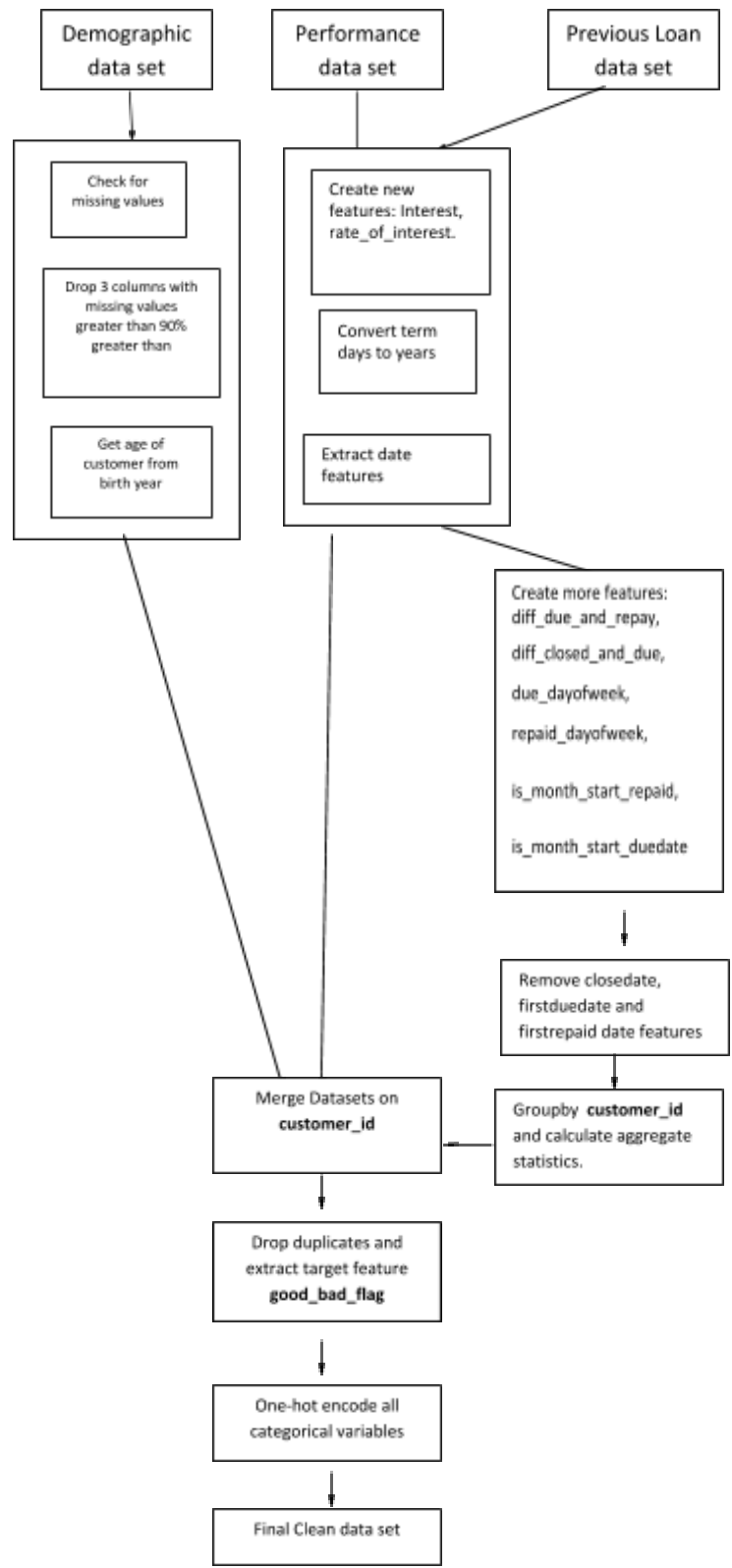

Fig, 1 Flowchart for data preprocessing

| Parameter | Value |
|---|---|
| n_estimators | 1000 |
| class_weight | binary |
| learning_rate | 0.01 |
| sub_sample | 0.8 |
| reg_alpha | 1 |
| reg_lambda | 1 |
| max_depth | 6 |

**Table 2.** Optimal parameters for our algorithm after Gridsearch

### 4.1 Confusion Matrix

The confusion matrix is an important 2 dimensional matrix that contains information about the actual classes and the predicted classes of a classifier. In the loan application data used in this paper, the number 0 represents the loan default category, and 1 represents the normal category. Table 3. shows the calculated confusion matrix of XGBoost the data.

| | Predicted Class | | |
|---|---|---|---|
| Actual Class | | Good | Bad |
| | Good | 175 | 15 |
| | Bad | 14 | 670 |

**Table 3** Confusion Matrix for XGBoost

### 4.2 Accuracy

Accuracy measures the proportion of correctly classified predictions; it is defined by the formula:

$$\text{Accuracy} = \frac{TP+TN}{TP+TN+FP+FN}$$

Where TP is the number of true positives, TN is the number of true negatives, FP is the number of false positives and FN is the number of false negatives.

### 4.3 Recall

Recall measures the ability of the classifier to predict correctly instances of a certain class; it is also called the TPR (true positive rate):

$$\text{Recall} = \frac{TP}{TP+FN}$$

### 4.4 Precision

Precision measures the proportion of predictions made by the classifier as positive that are actually positive:

$$\text{Precision} = \frac{TP}{TP+FP}$$

### 4.5 F1-score

F1-score is the harmonic average of precision and recall:

$$\text{F1-score} = \frac{2*Precision*Recall}{Precision+Recall}$$

| Metric | Score (%) |
|---|---|
| Accuracy | 79 |
| Precision | 97 |
| Recall | 79 |
| F1_score | 87 |

**Table 4.** Performance metric for XGBoost algorithm on the loan dataset.

### 4.6 ROC

ROC (receiver operating characteristics curve) is a visualization technique for showing a classifier's performance. It represents the sensitivity and specificity of the classifier. The ROC curve is a two-dimensional curve with the FPR (false positive rate) as the X-axis and the TPR (true positive rate) as the Y-axis. The ranges of the ROC curve runs from (0, 0) to (1, 1). To compare models, we calculate the area under the ROC curve (AUC). The larger the AUC, the better the model.

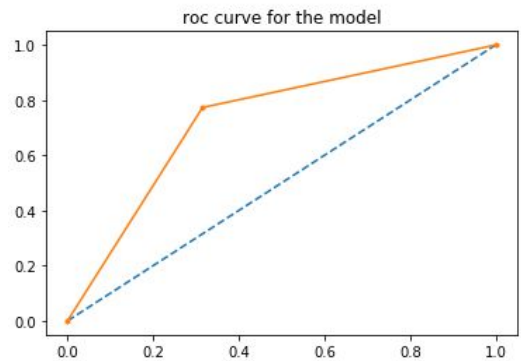

Fig. 2. ROC area curve of XGBoost on the loan data set.

### 4.7 Feature Importance

One of the important results of this study is to determine the important features that help the classifier to correctly predict loan default. This helps in business intelligence and decision making. Fig.3 shows the top 10 most important features. The location and age of the customer are the two most important features from the result of this plot.

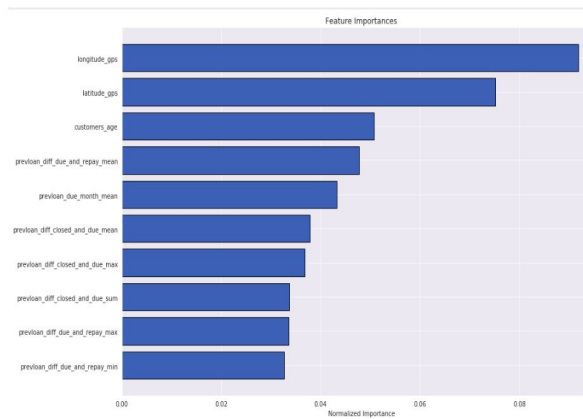

Fig. 3. ROC area curve of XGBoost on the loan data set.

## 5  Conclusion

In this paper, we have successfully used the Extreme Gradient Boosting (XGBoost) for bank loan default prediction. The task was to predict if a loan applicant will default in loan payment or not. The analysis was implemented in the python programming language, and performance metrics like accuracy, recall, precision, f1-score were calculated. From the analysis, we found out that the most important features used by our model for predicting if a customer would default in payment or not depends heavily on the location and age of the customer. This paper provides an effective basis for loan credit approval in order to identify risky customers from a large number of loan applicants using predictive modeling.